\documentclass[aps,prb,twocolumn]{revtex4}
\usepackage{bm,amsmath,graphicx}
\begin{document}

\title{Phase signal definition for electromagnetic waves in X-ray crystallography}

\author{S\'ergio L. Morelh\~ao}
\affiliation{Instituto de F\'{\i}sica, Universidade de S\~ao Paulo, CP 66318, 05315-970 S\~aoPaulo, SP, Brazil}

\begin{abstract} 
Diffracted X-ray waves in crystals have relative phases regarding the mathematical format used to describe them. A forward propagating wave can be defined with either negative or positive time evolution, i.e. ${\bm k}\cdot{\bm r} - \omega t$ or $\omega t - {\bm k}\cdot{\bm r}$. Physically measurable quantities are invariant with respect to the choice of definition. This fact has not been clearly emphasized neither extensively explored when deriving well-established equations currently being used in many X-ray diffraction related techniques. Here, the most important equations are generalized and consequences of conflicting undertaken definitions discussed.
\end{abstract}

\maketitle

\section{Introduction}

Electromagnetic waves are in general described by an expression such as 

\begin{equation}
{\bm D}({\bm r},t) = {\bm D}_0({\bm r}) e^{i S (\omega t - {\bm k}\cdot{\bm r})}
\label{planewave}
\end{equation}
where $S=\pm1$ stands for the global phase signal definition of monochromatic forward propagating waves. Both choices of phase signal ($+$ or $-$) are allowed and both can be founded in several theoretical approaches of the X-ray diffraction phenomenon in crystals.

Physically measurable quantities are invariant regarding two distinct and independent choices: one is the choice on how to describe the forward propagating X-ray wave, represented here by the options $S=-1$ and $S=+1$; and the other is the choice of origin for the spatial position ${\bm r}$. In the literature of X-ray diffraction (see for instance Refs.~\onlinecite{inte2001,auth2001,ashc1976,kitt1996,jens2001}) these two distinct choices have been of particular relevance for deriving important mathematical expressions, such as those of the structure factor of a crystal unit cell and of the atomic form factor. Whatever is the chosen option, $S=\pm1$, it can be compensated by the choice of origin so that the expressions are obtained in their standardized format without showing any dependence to the phase signal option. 

This standard procedure links the choice of origin to the choice of phase signal, which is unreal since both choices are independent from each other. The real fact is that the available expressions are in agreement with only one value of $S$ for each given choice of origin. Consequently, the propagating waves in the medium must be described according to the implicitly undertaken choices, otherwise measurable features such as the positioning of standing wavefields \textemdash formed inside crystals undergoing diffraction\textemdash~ as well as invariant phase values \textemdash accessible via $3$-beam diffraction experiments \cite{hart1961}\textemdash~ will be affected by conflicting definition of the global phase signal. 

\begin{figure}
\includegraphics[width=3.2in]{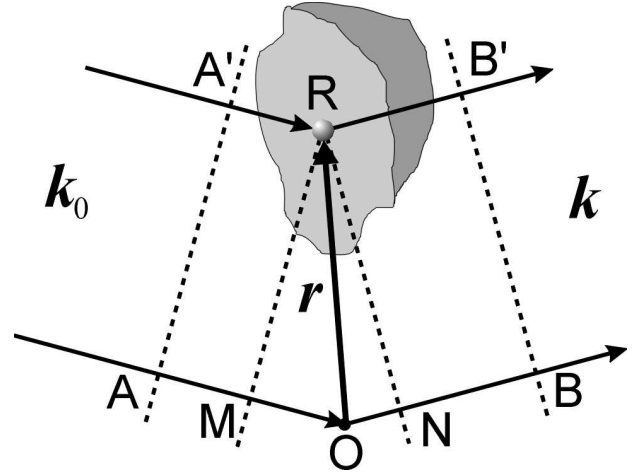}
\caption{The relative phase of the scattered wave from $R$, regarding an arbitrary choice of origin at $O$, is given in terms of incident ${\bm k}_0$ and scattered ${\bm k}$ wavevectors since $k\overline{MO} = -{\bm k}_0\cdot{\bm r}$ and $k\overline{ON} = {\bm k}\cdot{\bm r}$, as demonstrated in Eq.~(\ref{phasediff}). For sake of simplicity in analyzing the dependence with the global phase signal $S$, the origin is chosen very far from the charge-density distribution $\rho({\bm r})$, to asssure that $({\bm k}-{\bm k}_0)\cdot{\bm r}$ would have the same signal, in our case positive, for any point $R$ where $\rho({\bm r})\neq0$.}
\end{figure}

In this work the conflicting points are demonstrated and commonly used expressions in X-ray crystallography are generalized for any preference of signal choice $S$. Signal shifts owing to geometrical factors are avoided by choosing the origin further way from the diffracting volume. The consequences only on the choice of $S$ is then exploited and discussed in the rest of the paper. Moreover, to facilitate part of the demonstration, an adaptation of the Stokes relations for Bragg reflection in perfect low-absorbing crystals is presented.

\section{Susceptibilities to phase signal definition}

\subsection{Structure factor}

As depicted in Fig. 1, the scattered wave $$\Delta{\bm D}({\bm k})={\bm D}(B')+{\bm D}(B)$$ towards direction ${\bm k}$ depends on the relative phase between the rays from the origin $O$ and from the electron density at any given position ${\bm r}$, as for instance at $R$. In the wavefront $\overline{BB'}$, $${\bm D}(B') = \Delta{\bm D}_Re^{iS[\omega t - k(\overline{A'R}+\overline{RB'})]}$$ and $${\bm D}(B) = \Delta{\bm D}_Oe^{iS[\omega t - k(\overline{OA}+\overline{OB})]} = \Delta{\bm D}_Oe^{i\varphi}.$$ Since the origin has been chosen outside the electron density distribution $\rho({\bm r})$, there is no scattering at the origin, i.e. $\Delta{\bm D}_O=0$. Also, the origin was chosen in a such way to guarantee that $({\bm k}-{\bm k}_0)\cdot{\bm r}>0$ to all positions inside the volume where $\rho({\bm r})\neq0$, hence

$$k(\overline{A'R}+\overline{RB'}) = k(\overline{AO}+\overline{OB}) - k(\overline{MO}+\overline{ON})$$ 
and 

\begin{eqnarray}
\Delta{\bm D}({\bm k}) &=& \Delta{\bm D}_Re^{iSk(\overline{MO}+\overline{ON})}e^{i\varphi}
\nonumber \\
&=& \Delta{\bm D}_Re^{iS({\bm k}-{\bm k}_0)\cdot{\bm r}}e^{i\varphi}.
\label{phasediff}
\end{eqnarray}

The choice-of-origin phase $\varphi$, is common to all elements of volume ${\rm d}V$ at any instant of time, and can be conveniently chosen to provide $e^{i\varphi}=1$. If $\Delta{\bm D}_R = {\bm D}_e\rho({\bm r}){\rm d}V$ (${\bm D}_e$ is the scattering amplitude of a single electron), the total scattered wave by the volume V is 

\begin{equation}
{\bm D}({\bm k}) = {\bm D}_e\int_V\rho({\bm r})e^{iS({\bm k}-{\bm k}_0)\cdot{\bm r}}{\rm d}V.
\label{scatwave}
\end{equation}

In X-ray crystallography, the structure factor $F_H$ corresponds to the total scattered wave from an unit cell, which is ${\bm D}_c({\bm k}) = {\bm D}_eF_H/V_c$~($V_c$ is the unit cell volume). Therefore, it follows from Eq.~(\ref{scatwave}) that

\begin{equation}
F_H = \sum_n f_n e^{iS2\pi{\bm H}\cdot{\bm r}_n} = |F_H|e^{i\delta_H}
\label{sfactor}
\end{equation}
where $2\pi{\bm H}={\bm k}-{\bm k}_0$, ${\bm H}$ is the diffraction vector of reflection H, and $f_n$ is the atomic form factor \cite{inte2001} of the atom at the position ${\bm r}_n$ in the unit cell. 

\subsection{Electron density of the unit cell}

The step from Eq.~(\ref{scatwave}) to Eq.~(\ref{sfactor}) is better comprehended when written the electron density of the unit cell 

\begin{equation}
\rho_c({\bm r})=\frac{1}{V_c}\sum_n f_n \delta({\bm r}-{\bm r}_n)
\label{rhoc1eqn}
\end{equation}
in terms of Dirac $\delta$-functions

\begin{equation}
\delta({\bm r}-{\bm r}_n) = \sum_{\bm H} e^{\pm i 2\pi {\bm H}\cdot({\bm r}-{\bm r}_n)}.
\label{dirac}
\end{equation}
The sum in ${\bm H}$ stands for three sum on integer numbers $h$, $k$ and $l$, running from $-\infty$ to $+\infty$ since ${\bm H}\cdot{\bm r}_n=hx_n+ky_n+lz_n$ and $x_n$, $y_n$ and $z_n$ are the fractional coordinates of the atoms in the unit cell. $hkl$ is in fact the M\"uller index of a given reflection H.

The $\pm$ option in the exponent of Eq.~(\ref{dirac}) is not related to the choice $S$ but it can be conveniently used to obtain

\begin{equation}
\rho_c({\bm r})=\frac{1}{V_c}\sum_{\bm H}F_H e^{-iS2\pi{\bm H}\cdot{\bm r}}. 
\label{rhoc2eqn}
\end{equation}

Methods for experimental determination of $\rho_c({\rm r})$, or structure determination methods, are based on the above expression Eq.~(\ref{rhoc2eqn}), although it has been standardized for the $S=+1$ choice. 

\subsection{Standing waves}

The X-ray reflection coefficient of a crystal with $N$ planes (crystal thickness $Nd$, lattice period $d$) in symmetric Bragg reflection geometry can be written as

\begin{equation}
R_N(\theta)=|R_N(\theta)|e^{i(\delta_H+\Omega)}
\label{reflcoef}
\end{equation}
where $\theta$ is the rocking curve angle, i.e. the angle between the incident X-ray wave and the lattice planes. $\delta_H$ is the phase of the structure factor in Eq.~(\ref{sfactor}), and $\Omega$ is known as the dynamical phase \cite{auth2001} varying from the value $\Omega_{Left}$ at the left shoulder of the diffraction peak to the value $\Omega_{Right}$ at the right shoulder; both values will be given latter on.

To calculate the positioning of the standing waves during the rocking curve, the incident ${\bm D}_I$ and reflected ${\bm D}_R$ waves at a given depth $h$ are approximated to

\begin{equation}
{\bm D}_I=D_0(h)\hat{\bm v}_0\>e^{iS(\omega t - {\bm k}_0\cdot{\bm r})}
\label{eieqn}
\end{equation}
and

\begin{equation}
{\bm D}_R=R_N(\theta)D_0(h)\hat{\bm v}\>e^{iS(\omega t - {\bm k}\cdot{\bm r})}
\label{ereqn}
\end{equation}
where $\hat{\bm v}_0$ and $\hat{\bm v}$ are their oscillation directions. A comparable reflection coefficient at all depths is assumed, i.e. $R_N(\theta)$ do not depend on $h$. All dependence with depth is accounted for $D_0(h)$, which is a smooth function with very small variation over the lattice period.

The standing wavefield is therefore ${\bm D}_{SW}={\bm D}_I + {\bm D}_R$ whose time-average intensity at any depth is proportional to

\begin{equation}
I_{SW} = \frac{|{\bm D}_{SW}|^2}{D_0^2} =
1+|R_N(\theta)|^2+2|R_N(\theta)|\hat{\bm v}_0\cdot\hat{\bm v}\cos\Phi,
\label{swint}
\end{equation}
providing antinodes (maxima of intensity) at 
\begin{equation}
\Phi = \delta_H + \Omega - 2\pi S {\bm H}\cdot{\bm r} = 2m\pi
\label{phieqn}
\end{equation}
for every integer value of $m$.

For sake of simplicity, a single atomic layer per lattice plane period is assumed so that ${\bm H}={\hat z}/d$, $$ F_H = e^{iS 2\pi z_0/d}\sum_n f_n~,$$ and $\delta_H = S 2\pi z_0/d$ where $z_0$ gives the atomic layer position in the crystal unit cell. If $z_0=0$, the unit cell origin would fall on top of the atomic layer.

From Eq.~(\ref{phieqn}) we have the antinodes at 

\begin{equation}
z_A(\Omega) = z_0 + Sd(\Omega/2\pi - m)
\label{znode}
\end{equation}
as a function of the dynamical phase $\Omega$, and 

\begin{equation}
\Delta z_A = z_A(\Omega_{Right}) - z_A(\Omega_{Left}) = S\frac{\Delta\Omega}{\pi}\frac{d}{2}
\label{deltaz}
\end{equation}
as the displacement of the antinodes during the crystal's rocking curve. Such displacement must be invariant regarding the choice of signal $S$ and,
therefore, it can be assumed that

\begin{equation}
\Delta\Omega = \Omega_{Right} - \Omega_{Left} = -S\pi
\label{deltaomega}
\end{equation}
since the displacement is experimentally observed \cite{truc1976, bedz1984} as been $\Delta z_A=-d/2$.

It is widely known from dynamical diffraction theory \cite{ auth2001} that $\Omega_{Left} = \pi$ and $\Omega_{Right} = 0$. However, a clear statement should be made to emphasize that this phase variation is relative to the global phase signal $S=+1$, as demonstrated below.

\subsection{Dynamical phase shift versus global phase signal}

\begin{figure}
\includegraphics[width=3.2in]{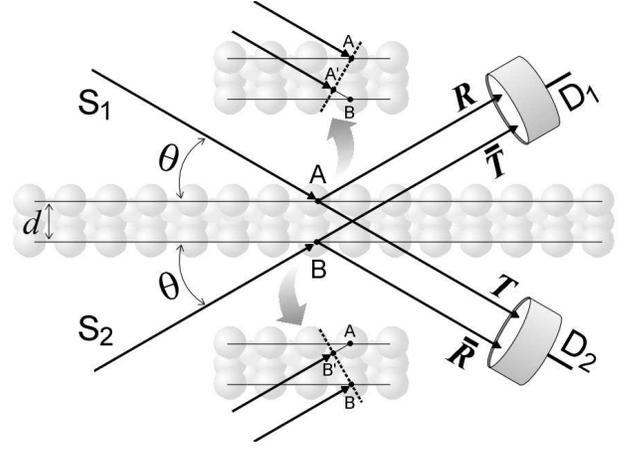}
\caption{Reflection $R$ and $\bar{R}$, and transmission $T$ and $\bar{T}$, coefficients for electromagnetic plane-waves in very thin and uniform planes of low-absorbing matter. Stokes relations for thin film optics \cite{knit1976} provide conservation of the counting rate $|R + \bar{T}|^2$ plus $|\bar{R} + T|^2$, at detectors $D_1$ and $D_2$ respectively, regarding the total impinging intensity from the identical sources $S_1$ and $S_2$. The top and bottom insets illustrate the phase delay $\varphi=-Sk\overline{A'B}=-Sk\overline{B'A}$ of the transmitted waves across the plane thickness.}
\end{figure}

Coherent scattering and transmission of electromagnetic waves by very thin and uniform planes of low-absorbing matter are describable by reflection, $R$ and $\bar{R}$, and transmission, $T$ and $\bar{T}$, coefficients as depicted in Fig.~2. The condition

\begin{equation}
R\bar{T}^{*}+R^{*}\bar{T}+\bar{R}T^{*}+\bar{R}^{*}T=0
\label{bse1}
\end{equation}
is required by energy conservation, which stipulates phase relationships between the reflected and transmitted waves, as usually obtained for laser beam splitters \cite{loud2000} or, equivalently, by the Stokes relations for time reversibility of wave's propagation \cite{knit1976}.

Without loosing generality, these coefficients can be written as

\begin{eqnarray}
R&=&\pm i|R|\>e^{i(\delta+\bar{\varphi})},~~T=|T|\>e^{i\varphi},\nonumber\\
\bar{R}&=&\pm i|\bar{R}|\>e^{i(\bar{\delta}+\varphi)},~~{\rm and}~~\bar{T}=|\bar{T}|\>e^{i\bar{\varphi}}
\label{phaeqn}
\end{eqnarray}
in order to fulfill Eq.~(\ref{bse1}). $\delta\pm90^{\circ}$ and $\bar{\delta}\pm90^{\circ}$ are the amount by which the phases of the reflected waves can differ from the phases of the transmitted ones, and $\varphi$ and $\bar{\varphi}$ are the phase delays across the plane, whose thickness $d$ is comparable to the wavelength $\lambda$. 

To determine the phase delays consider first only the source $S_1$ in Fig. 2. As shown in the top inset, if ${\bm D}(A)$ is the incident wave at point $A$, ${\bm D}(B)=T{\bm D}(A)$ is the transmitted wave at point $B$ passing through point $A'$. Then, $$|T|{\bm D}_0e^{iS(\omega t - {\bm k}_0\cdot{\bm r}_B)}=T{\bm D}_0e^{iS(\omega t - {\bm k}_0\cdot{\bm r}_{A'})}$$ where ${\bm k}_0\cdot{\bm r}_{A'}={\bm k}_0\cdot{\bm r}_A$ since ${\bm r}_A$ and ${\bm r}_{A'}$ stand for positions on the same wavefront $\overline{AA'}$, providing $$T = |T|\>e^{-iS({\bm r}_B-{\bm r}_{A'})\cdot{\bm k}_0}=|T|\>e^{-iSk\overline{A'B}}=|T|\>e^{i\varphi}.$$ Since the incidence angle $\theta$ is the same for both sources $\overline{A'B}~=~\overline{B'A}$ and, consequently, 

\begin{figure*}
\includegraphics[width=6.4in]{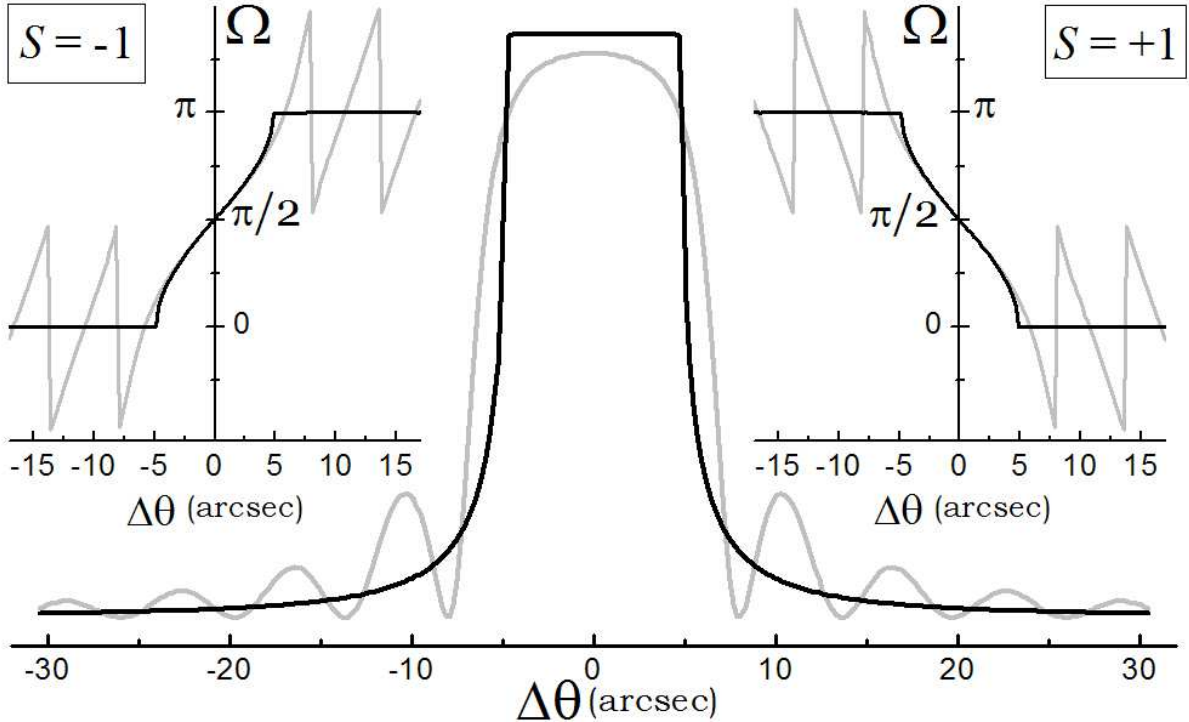}
\caption{Simulated X-ray reflectivity curves $|R(\theta)|^2$, for crystal thickness $Nd=2.5 \mu m$ (gray curve) and $Nd=21 mm$ (black curve, flat-top maximum equal to 1). The shift of the dynamical phase $\Omega$, Eq.~(\ref{reflcoef}), is shown on both sides of the rocking curves for the two possible choices of global phase signal in Eq.~(\ref{phdeqn}), i.e. $S=\pm1$. $\Delta\theta=\theta-\theta_{Bragg}$, $2d\sin\theta_{Bragg}=\lambda=1.54$\AA~(Bragg's Law), $d=3.14$\AA, $|R_1|=|\bar{R}_1| \simeq 3.0 \times10^{-4}$ (nearly the value for the silicon 111 reflection) and $\delta_H=0$. For simulation purposes some absorption had to be considered in the thick crystal case, then a linear absorption coefficient $\mu = a\sin\theta_{Bragg}/d \simeq 1 cm^{-1}$ was used. $a$ is defined in Eq.~(\ref{tlpeqn}).}
\end{figure*}

\begin{equation}
\varphi = \bar{\varphi}=-\>S\frac{2\pi}{\lambda}d\sin\theta\>.
\label{phdeqn}
\end{equation}

The X-ray reflectivity of a crystal, as given in Eq.~(\ref{reflcoef}), is obtained by stacking lattice planes whose reflection and transmission coefficients are written in the same format used above, Eq.~(\ref{phaeqn}). Stacking in geometrical progression of $N=2^n$ planes ($n = 1, 2,...\>$) is the fastest way to go from a single lattice plane to the desired thickness $Nd$. It is possible by means of the recursive equation

\begin{subequations}
\begin{equation}
\left[\begin{array}{c} 
R_N\\ 
\bar{R}_N
\end{array}\right]=
\left(1+\frac{T_{N/2}\bar{T}_{N/2}}{1-\bar{R}_{N/2}R_{N/2}}\right)\left[\begin{array}{c}
R_{N/2}\\
\bar{R}_{N/2}
\end{array}\right] 
\label{rneqn}
\end{equation}
and
\begin{equation}
\left[\begin{array}{c} 
T_N\\ 
\bar{T}_N
\end{array}\right] = \frac{1}{1-R_{N/2}\bar{R}_{N/2}}\left[\begin{array}{c}
T_{N/2}^2\\
\bar{T}_{N/2}^2
\end{array}\right] 
\label{tneqn}
\end{equation}
\label{rntneqn}
\end{subequations}
whose derivation is very similar to the Airy's formula of the Fabry-Perot interferometer, and where 

\begin{subequations}
\begin{equation}
\left[\begin{array}{c} 
R_1\\ 
\bar{R}_1
\end{array}\right]=-i\frac{r_e\lambda|C|d}{V_c\sin\theta}
\left[\begin{array}{c}
F_H\\
F_{\bar{H}}
\end{array}\right]e^{i\varphi} 
\label{rlpeqn}
\end{equation}
and
\begin{equation}
\left[\begin{array}{c} 
T_1\\ 
\bar{T}_1
\end{array}\right]=\left[\begin{array}{c}
(1-|R_1|^2-a)^{1/2}\\
(1-|\bar{R}_1|^2-a)^{1/2}
\end{array}\right]e^{i\varphi}\>. 
\label{tlpeqn}
\end{equation}
\end{subequations}
$C = \hat{\bm v}_0\cdot\hat{\bm v}$, $\bar{\bm H}=-{\bm H}$, $r_e=2.818\times10^{-5}$\AA ~(classical electron radius), and $a$ stands for photoabsorption probability on each individual lattice plane, as discussed in details and compared to other diffraction theories elsewhere \cite{more2005}.

Fig.~3 shows the behavior of the dynamical phase across the reflectivity curve as a function of the phase signal $S$ in Eq.~(\ref{phdeqn}). As expected, $\Omega_{Left} = \pi$ and $\Omega_{Right} = 0$ for the $S=+1$ choice, but $\Omega_{Left} = 0$ and $\Omega_{Right} = \pi$ for the other choice, i.e. $S=-1$, which is also in agreement with Eq.~(\ref{deltaomega}).

\section{Discussion on phase invariants}

Phase determination is a fundamental problem in X-ray crystallography. Reflection intensities only provide $|F_H|$ as experimental data input for $\rho_c({\rm r})$ in Eq.~(\ref{rhoc2eqn}). Then, what phase $\delta_H$ should be assigned to the structure factor $F_H = |F_H|\exp(i\delta_H)$ of each crystal reflection? 

For decades, great effort has been dedicated in developing and improving methods to estimate reflection phases, but since these are relative choice-of-origin values, the estimable quantities are in fact the differences among reflection phases, better known as phase invariants \cite{giac1999}. For instance, consider three reflections $G$, $H$ and $L$ whose diffraction vectors fulfill the condition ${\bm G}={\bm H}+{\bm L}$. Hence, in the structure factor ratio $$\frac{F_HF_L}{F_G}=\frac{|F_H||F_L|}{|F_G|}e^{i\Psi},$$ the triple phase $\Psi = \delta_H+\delta_L-\delta_G$ is invariant regarding the choice of origin in Fig.~1.

Three-beam diffraction experiments ~\cite{hart1961,shen1987,more2002,more2005b} are sensitive to $\Psi$. In azimuthal scan mode, the diffracted intensity of reflection $G$ is modulated by the excitation of the reflection $H$ during the crystal rotation $\phi$ around ${\bm G}$. This intensity modulation is approximately given by

\begin{eqnarray}
{\texttt I}(\phi) = |{\bm D}_G|^2 + |{\bm D}_{HL}(\phi)|^2 + & \nonumber \\
+ 2|{\bm D}_G\cdot{\bm D}_{HL}(\phi)|&\cos(\Psi + \Omega),
\label{tbint}
\end{eqnarray}
which explicitly depends on the dynamical phase $\Omega$ of the reflection $H$. ${\bm D}_G$ and ${\bm D}_{HL}(\phi)$ stand for the diffracted wavefields from reflection $G$, kept constant during the $\phi$-scan, and from the detour reflection $H+L$. 

Although the signal of $\Psi+\Omega$ changes with $S$, the invariance of ${\texttt I}(\phi)$ is assured by the cosine, an even function. However, both angles must stand for the same choice of $S$ in Eqs.~(\ref{sfactor}) and (\ref{phdeqn}). Otherwise, the experimental determination of $\Psi$ by $\phi$-scan data analysis would be compromised.

Besides the signal of $\Psi$, its modulus will also change with $S$ when the $f'$ and $f''$
corrections \cite{inte2001} of the atomic form factor \textemdash the so-called anomalous
scattering corrections\textemdash~ are taken into account. According to the structure factor
expression in Eq.~(\ref{sfactor}) 
\begin{equation}
|F_H(S)|e^{i\delta_H(S)} = \sum_n (f_0 + f' + if'')_n e^{iS2\pi{\bm H}\cdot{\bm r}_n},
\label{sf=sf}
\end{equation}
and then $|F_H(+1)|\neq|F_H(-1)|$ and $\delta_H(+1) \neq -\delta_H(-1)$ when $f''\neq0$. It is
unacceptable since these inequalities would imply in different diffracted beam intensities for each
possible choice of wave representation, $S=+1$ or $S=-1$. To avoid such artificial fact, the $f''$
correction must bear its depence with $S$, so that
\begin{equation}
f = f_0 + f' + iSf''.
\label{affeqn}
\end{equation}

\section{Conclusions}

Structure factor, dynamical phase shift, phase invariants, and complex atomic form factor are values susceptible to the choice of global phase signal, as explicitly demonstrated here. When a measurable quantity depends on more than one of these values, they must be calculated for the same choice of phase signal.  

Recursive equations based on Stokes relations are applied for calculating the reflectivity of low-absorbing crystals with a finite number of lattice planes. These recursive equations are extremely simple and, yet, capable of describing important features of the diffraction phenomenon such as kinematical diffraction (thin crystals), primary extinction (maximum reflectivity equal to 1), intrinsic width of Bragg reflections in either kinematical or dynamical diffraction regimes, and the phase shift of the diffracted waves across the reflection domain.

\begin{acknowledgments} 
The author would like to thank Prof. Paulo A. Nussenzveig for valuable discussions and very kind revision of the manuscript, as well as the Brazilian founding agencies FAPESP, grant number 02/10387-5, and CNPq, proc. number 301617/95-3.
\end{acknowledgments}

\end{document}